\documentclass[]{elsart}

\usepackage{graphicx}
\usepackage{amssymb}

\begin{document}
\begin{frontmatter}


\journal{SCES'2001}


\title{Temperature Dependent $5f$-states in URu$_2$Si$_2$}

%
%
%
%
%
%

\author[LBNL]{J. D. Denlinger\corauthref{1}}
\author[UM]{G.-H. Gweon}
\author[UM]{J. W. Allen}
\author[LANL]{J. L. Sarrao}

%
 
\address[LBNL]{Advanced Light Source, Lawrence Berkeley Nat'l
Lab, Berkeley, CA 94720, USA}
\address[UM]{Randall Lab. of Physics, Univ. of Michigan, Ann Arbor,
MI 48109-1120, USA}
\address[LANL]{Los Alamos National Laboratory, Los Alamos, NM 87545, USA}

%
%
%
%


%
%
%
%

\corauth[1]{Corresponding Author: MS 7-222, Lawrence Berkeley National
Laboratory, 1 Cyclotron Road, Berkeley, CA 94720, USA.  Phone: (510)
486-5648, Fax: (510) 495-2067, Email: JDDenlinger@lbl.gov}


\def\EF{$E_{\rm F}$}
\def\RuSi {Ru$_2$Si$_2$}

\begin{abstract}

A dramatic temperature dependent enhancement of U 5$f$ spectral weight at \EF\
is observed in angle-resolved photoemission measurements of U\RuSi\ at the
center of an X-point hole-pocket.  Comparison of this temperature dependent
behavior for excitation both at and below the U $5d$$\rightarrow$$5f$ resonant
threshold is presented.
\end{abstract}

%
%



\end{frontmatter}

%
%
%
%
%

\def\EF{$E_{\rm F}$}
\def\RuSi {Ru$_2$Si$_2$}
\def\kv {$\bf k$}
\def\kpar {$k_{\parallel}$}
\def\TK {$T_{\rm K}$}
\def\deg {$^{\circ}$}
\def\invang {\AA$^{-1}$}

A fundamental property of Kondo systems is a temperature scale below which local
magnetic moments are gradually quenched as the system settles to a singlet ground
state.  An important aspect of the Anderson single impurity and lattice models for
photoemission spectroscopy of $f$-electron systems is the prediction of an
observable temperature dependence of the associated Kondo resonance near \EF\ as
the system is cooled through the Kondo temperature \TK.  For impurity models,
either a sharpening or an increase of the $k$-integrated photoemission $f$-spectral
weight is expected for Ce or Yb, respectively, with the difference arising from the
former having one $f$-electron and the latter one $f$-hole \cite{Gunnarsson87}. 
Such temperature dependence has been reported for Ce \cite{Garnier97} and Yb
\cite{Tjeng93} compounds, but also disputed \cite{Arko99}.  Predictions for lattice
models exist \cite{Tahvildar98}, but thus far the models are not realistic as to
the local orbital degeneracy or conduction electron number for Ce, Yb or U. 
Experimentally no significant temperature variation in photoemission has yet been
reported for any U compound \cite{Arko99}.

U\RuSi, with an intermediate Kondo temperature of $\approx$70 K \cite{Bonn88},
is favorable for such a temperature-dependent study of the $f$-weight at
different locations in\kv-space.  Previous ARPES experiments on U\RuSi\
\cite{Denlinger00,Denlinger01} have (a) determined the crystal inner potential
($V_0\approx12$ eV), (b) mapped the basic $d$-band structure along high symmetry
directions and (c) established the existence of hole pockets at the $\Gamma$, Z and
X-points of the Brillouin zone, and (d) provided k-dependent $5f$ spectral
signatures of the Anderson lattice using resonant ARPES.  In this paper, a dramatic
temperature-dependence of the U
$5f$ spectral weight at the center of an X-point hole-pocket is presented and
compared at on- and off-resonance photon energies.  Previously no significant
temperature variation of angle-integrated valence spectra from scraped single
crystals of U\RuSi\ \cite{Yang96} was observed. 

U\RuSi\ has the ThCr$_2$Si$_2$ crystal structure with a body-centered
tetragonal Brillouin zone.  Single crystal U\RuSi\ samples were cleaved in
ultra-high vacuum ($\leq$5$\times$10$^{-11}$ torr) at 100 K exposing the [001]
surface for ARPES measurements at ALS Beamline 10.0.1.  Temperature measurements
and regulation of a flowing He cryostat was performed with Si diode sensors
attached close to the base of transferable sample stubs.  Photon energies above
and below the U $5d\rightarrow5f$ absorption thresholds (108 eV) were used to
compare spectra dominated by $d$-band spectral weight to spectra with U $5f$ weight
resonantly enhanced.  A total instrumental resolution of $\approx$40 meV and full
angular acceptance of $\approx$0.36\deg\ was employed.  

Fig. 1 shows such angular-resolved spectra measured at the X-point
with both on- and off-resonance photon energies for temperatures in the
experimental sequence of 100 K, 50 K and 25 K.  Insets in Fig. 1 illustrate
the existence of a distinct hole-pocket centered on the X-point (102 eV) and
the photon energy dependence of the spectral weight.   Namely, the on-resonance
spectra (h$\nu$=108 eV) with enhanced $5f$ cross section, shows a clear
confinement of the U $5f$ spectra weight to the interior of the $d$-band
hole-pocket.  This behavior represents a basic signature in the
Anderson lattice model of $f$-$d$ mixing near an idealized $d$-band crossing
\cite{Tahvildar98,Denlinger01}.  Important for this
temperature dependent study is the relative absence of $d$-states below \EF\ at
the center of this hole-pocket, thus providing a nearly ideal location at which
$f$-states alone can be probed with minimal interference from other spectral
weight.  

In the spectra of Fig. 1, a striking additional enhancement of the \EF\ peak is
observed for lower temperatures.  Using 100K spectra taken just outside the X-point
hole pocket at \kpar=1.5 \invang, as a model for the temperature-invariant
background weight (short dashed spectrum in Fig. 1), the 5$f$ spectral weight
amplitude (area) enhancement at 108 eV, after subtraction of the background
spectrum, is approximately 2.2 (1.5).  For the off-resonance photon energy the 5$f$
amplitude (area) enhancement of 2.6 (2.1) is even more dramatic.  

A noticeable difference between the on- and off-resonance spectra is the
existence of some additional temperature-invariant on-resonance spectral weight 
between 0.1 and 0.3 eV below \EF. This 5$f$ spectral weight is approximately
isolated by taking the on- minus off-resonance difference of the spectra outside the
X-point pocket (see long dashed spectrum in Fig. 1(a)).  Probing the spatial
homogeneity of the sample surface reveals that this spectral weight is enhanced at
``bad'' regions of the sample surface where the observation of $d$-band dispersions
is also weakened. These locations are approximately correlated to macroscopically
rough topography  regions with enhanced surface area and possibly faster
oxidization. 

It is significant that angle-integrated spectra from this cleaved surface
do not reveal a discernable temperature dependence, due to the small ratio
of temperature-dependent to temperature-invariant spectral weight.  Hence, three
factors of $(i)$ reduced surface component, $(ii)$ angle-resolution, and $(iii)$
special \kv-space location are found to be necessary ingredients for this first
ever observation of temperature dependence in photoemission of a U compound,
providing an explanation for the previous null result for angle-integrated
photoemission of scraped single crystals of U\RuSi\ \cite{Yang96}. 

The above interpretation assumes that the observed temperature dependent spectral
weight impinging on \EF\ arises from the bulk \cite{surface}. The temperature range
of these variations is consistent with the bulk-sensitive observation by
point-contact spectroscopy of a resonance at \EF\ appearing below 60 K
\cite{Rodrigo97}.  Hence we are encouraged to interpret this large temperature
variation of U $5f$ spectral weight in ARPES of U\RuSi\ as possible evidence for
Kondo singlet condensation.  More detailed temperature dependent measurements are
in progress. 

Supported by the U.S. NSF at U. Mich. (DMR-9971611) and by 
the U.S. DoE at U. Mich. (DE-FG02-90ER45416) and at the Advanced Light Source
(DE-AC03-76SF00098).

%
%
%
%

%

\begin{figure}
     \centering
     \includegraphics{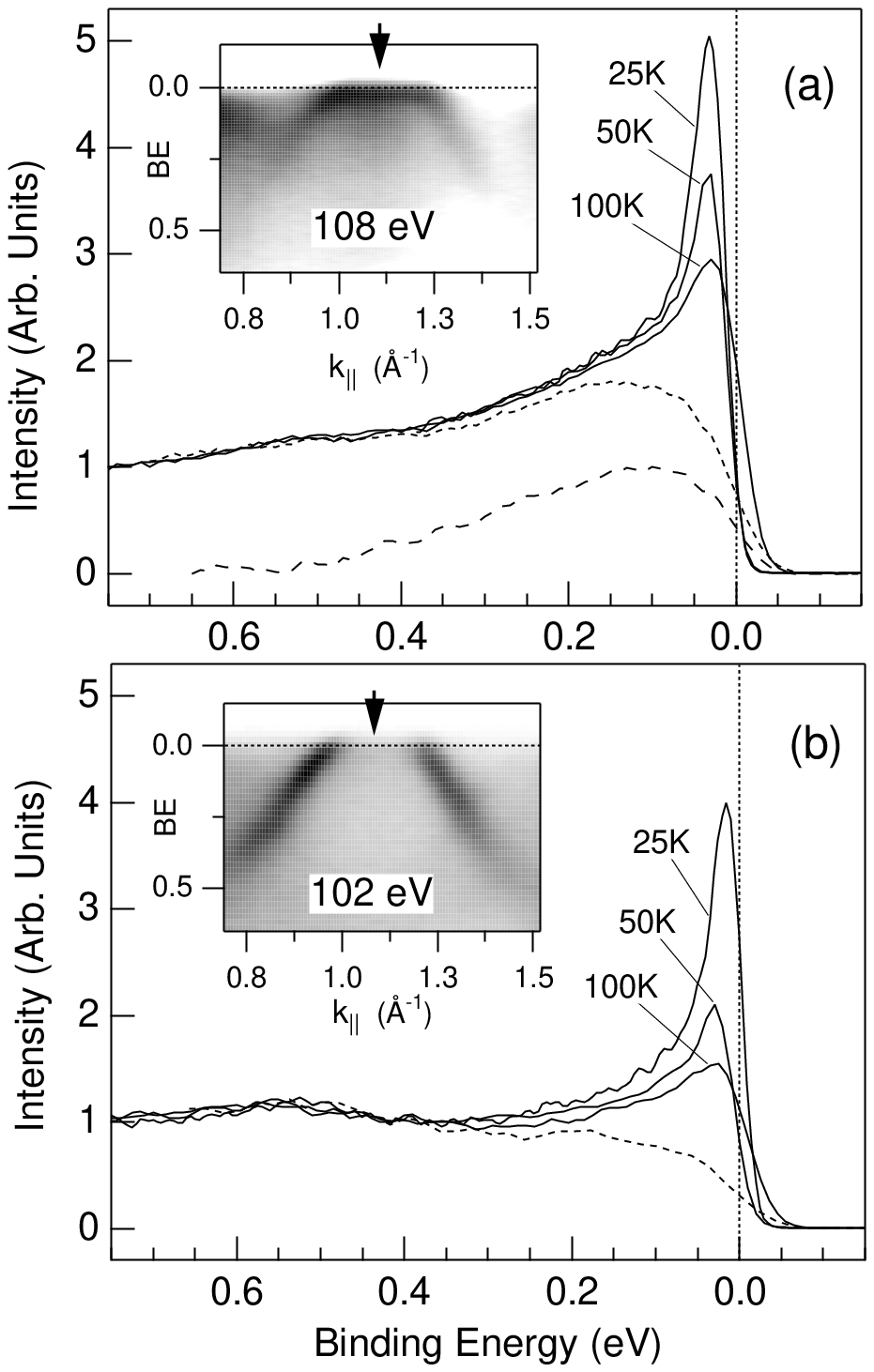}
     \caption{Temperature dependence of angle-resolved valence spectra at (a)
108 eV and (b) 102 eV for U\RuSi\ at the center of the X-point hole-pocket.
Insets show the corresponding X-point valence band structure at 100 K.  See
text for explanation of dashed-line spectra. } 
\end{figure}


\end{document}